\documentclass[12pt]{article}
\usepackage[utf8]{inputenc}
\usepackage{authblk}
\usepackage{setspace}
\usepackage[margin=1.25in]{geometry}
\usepackage{graphicx}
\usepackage{subcaption}
\usepackage{amsmath}
\usepackage{lineno}
\usepackage{lipsum}
\linenumbers
\usepackage[style=nejm, 
citestyle=numeric-comp,
sorting=none]{biblatex}
\usepackage{graphicx}
\usepackage{bm}
\usepackage{txfonts}
\usepackage{hyperref}
\date{\today}
\newcommand{\be}{\begin{eqnarray}}
	\newcommand{\ee}{\end{eqnarray}}

\newcommand{\bfp}{{\bf p}_{\perp}}

\usepackage{xcolor}
\begin{document}
	\nolinenumbers
\title{	Investigation on the higher twist TMD $h_3$ for proton in the light-front quark-diquark model}
\author[1]{Shubham Sharma\textsuperscript{$*$}}
\author[1]{Harleen Dahiya}
\affil[1]{Department of Physics, Dr. B. R. Ambedkar National Institute of Technology, Jalandhar 144027, India}
\affil[*]{s.sharma.hep@gmail.com}
%

\onehalfspacing
\maketitle

\date{}

\begin{abstract}
The higher twist T-even transverse momentum dependent distribution (TMD) \\$h_3(x, {\bf p_\perp^2})$ for the proton has been examined in the light-front quark-diquark model (LFQDM). By deciphering the unintegrated quark-quark correlator for semi-inclusive deep inelastic scattering (SIDIS), we have derived explicit equations of the TMD for both the scenarios when the diquark is a scalar or a vector. Average as well as average square transverse momenta have been computed for this TMD. Additionally, we have discussed its transverse momentum dependent parton distribution function (TMDPDF) $h_3(x)$.

\end{abstract}

\newpage
\section{Introduction}
Physics of hadrons includes its tomography in partonic degrees of freedom. This is achieved theoretically by 3-dimensions functions like transverse momentum dependent parton distributions (TMDs) and generalized parton distributions (GPDs). TMDs encode momentum information of the parton along both longitudinal and transverse directions and is linked experimentally to semi-inclusive deep inclusive scattering (SIDIS). Higher twist TMDs have been an arising topic of interest \cite{SHARMA2023116247}. In the present work, we have studied the twist-4 TMD $h_3(x,{\bf p_\perp^2})$.
\section{Light-Front Quark-Diquark Model (LFQDM) \label{secmodel}}
We examine the present problem by using the LFQDM where the spin-flavor $SU(4)$ structure of proton has been stated as a composite of isoscalar-scalar diquark singlet $|u~ S^0\rangle$, isoscalar-vector diquark $|u~ A^0\rangle$ and isovector-vector diquark $|d~ A^1\rangle$ states \cite{Maji:2016yqo}. 
The general form of light-front wave functions (LFWFs) $\varphi^{(\nu)}_{i}(x,\bfp)$ is derived from the soft-wall AdS/QCD prediction \cite{Maji:2016yqo}. The parameters of the model have been given in Ref. \cite{Maji:2016yqo}.

\section{Quark Correlator and Parameterization}
 The unintegrated quark-quark correlator for SIDIS is defined as \cite{Goeke:2005hb}
\begin{eqnarray}
	\Phi^{\nu [\Gamma]}(x,\textbf{p}_{\perp};S)&=&\frac{1}{2}\int \frac{dz^- d^2z_T}{2(2\pi)^3} e^{ip.z} \langle P; S_f|\overline{\psi}^\nu (0)\Gamma \mathcal{W}_{[0,z]} \psi^\nu (z) |P;S_i\rangle\Bigg|_{z^+=0}. 	\label{cor}
\end{eqnarray}
The momentum of proton, quark and diquark is $P\equiv\big(P^+,\frac{M^2}{P^+},\textbf{0}_\perp\big)$, $p \equiv \big(xP^+, \frac{p^2+|\bfp|^2}{xP^+},\bfp \big)$ and $P_X \equiv \big((1-x)P^+,P^-_X,-\bfp\big)$ respectively. 
The value of Wilson line $\mathcal{W}_{[0,z]}$ is chosen to be $1$. The quark-quark correlator has been parameterized for the Dirac matrix structure  
$\Gamma=\bm{i}\sigma^{i-}\gamma_5$ as \cite{Goeke:2005hb}
	\begin{eqnarray}
	\hspace{-5mm}    
	\Phi^{\nu[\bm{i}\sigma^{i-}\gamma_5]}
	&=&\frac{M^2}{(P^+)^2}[\bm{S}_{T}^i{\color{teal} \bf h_3}+\lambda\frac{\bm{p}_{T}^i}{M}h_{3L}^{\perp}+\frac{(\bm{p}_T^i\bm{p}_{T}^j-\frac{1}{2}\bm{p}_T^2g_T^{ij})\bm{S}_{Tj}}{M^2}{ h_{3T}^{\perp}}-\frac{\epsilon_T^{ij}\bm{p}_{Tj}}{M}h_3^{\perp}]. 
	   \label{eqtmdlist4}                        
\end{eqnarray}
\section{Result and Discussion}
\begin{figure*}
	\centering
	\begin{minipage}[c]{0.98\textwidth}
		(a)\includegraphics[width=7.0cm]{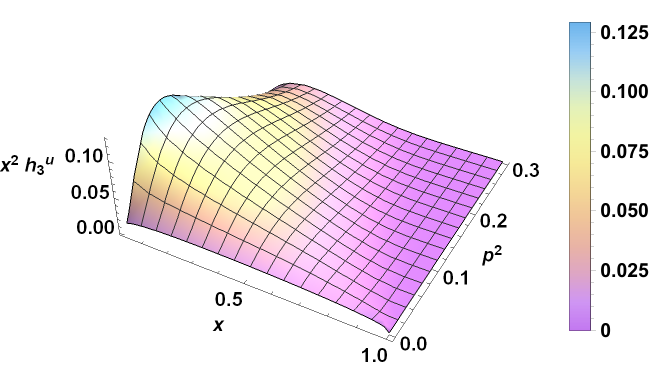}
		\hspace{0.05cm}
		(b)\includegraphics[width=7.0cm]{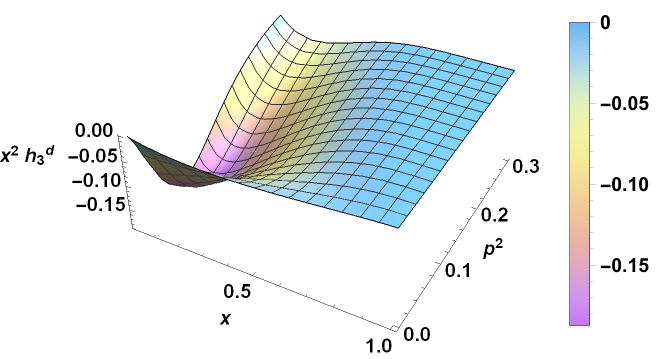}
		\hspace{0.05cm}
	\end{minipage}
	\caption{\label{fig3d} The TMD $x^2 h_{3}^{\nu}(x, {\bf p_\perp^2})$ is plotted with respect to $x$ and ${\bf p_\perp^2}$.}
\end{figure*}

After solving Eq. (\ref{cor}) and (\ref{eqtmdlist4}) for TMD $h_3^{\nu}(x, {\bf p_\perp^2})$, we get
\begin{eqnarray}
	x^2~h_{3}^{\nu}(x, {\bf p_\perp^2}) &=& \frac{1}{16 \pi^3} \Bigg({C_{S}^{2} N_s^2}-\frac{1}{3} {{C_{A}^{2}} |N_0^{\nu}|^2})\Bigg)\bigg[\frac{m}{M}|\varphi_1^{\nu}|^2 + \frac{\bfp^2}{M^2 x}|\varphi_2^{\nu}|\bigg]^2.
\end{eqnarray}
In Fig. \ref{fig3d}, the TMD $x^2 h_{3}^{\nu}(x, {\bf p_\perp^2})$ has been plotted for both the $u$ and $d$ quarks. It has been observed that the magnitude of TMD $x^2 h_{3}^{\nu}(x, {\bf p_\perp^2})$ is maximum at $x \sim 0.1$, $p_\perp^2 \sim 0.08~\mathrm{GeV}^2$ and it decreases as we move away from it. The maximum possibility of obtaining the helicity combination of proton corresponding to $x^2 h_{3}^{\nu}(x, {\bf p_\perp^2})$ is when the quark carry $10$ to $15\%$ of proton's longitudinal momenta. The value of average transverse momentum $\langle p_\perp \rangle$ and the average square transverse momentum $\langle p_\perp^2 \rangle$ of TMD $h_{3}^{u}(x, {\bf p_\perp^2})$ for $u (d)$ quarks computed comes to be $0.26~\mathrm{GeV}$ $(0.27~\mathrm{GeV})$ and $0.07~\mathrm{GeV}^2$ $(0.08~\mathrm{GeV}^2)$ respectively. The trend of TMDPDF $x^2 h_{3}^{\nu}(x)$ is identical to TMD plot at $0.2~\mathrm{GeV}^2$.


\printbibliography

\end{document}